\pdfoutput=1

\documentclass[fleqn,usenatbib]{mnras}

\usepackage{newtxtext,newtxmath}

\usepackage[T1]{fontenc}
\usepackage{ae,aecompl}

\usepackage{graphicx}	
\usepackage{amsmath}	
\usepackage{amssymb}	

\newlength{\VSpaceBeforeTabBib}
\newlength{\VSpaceBeforeTabFoot}
\usepackage{natbib}

\newcommand{\orcid}[1]{\href{https://orcid.org/#1}{\textcolor[HTML]{A6CE39}{\aiOrcid}}}
\hypersetup{draft}

\title[Proxima Centauri - \emph{AstroSat} view]{Proxima Centauri - the nearest planet host observed simultaneously with  \emph{AstroSat}, \emph{Chandra} and \emph{HST} }

\author[S. Lalitha et al.]{S. Lalitha$^{1}$\thanks{E-mail: sairaml@bham.ac.uk},
J.H.M.M. Schmitt$^{2}$, 
K.P. Singh$^{3}$, 
P. C. Schneider$^{2}$, 
R. O. Parke Loyd$^{4}$, 
\newauthor K. France$^{5}$, 
P. Predehl$^{6}$,
 V. Burwitz$^{6}$ 
and J. Robrade$^{2}$
\\
$^{1}$School of Physics \& Astronomy, University of Birmingham, Edgbaston, Birmingham B15 2TT, UK\\
$^{2}$ Hamburger Sternwarte, University of Hamburg, Gojenbergsweg 112, 21029 Hamburg, Germany\\
$^{3}$Indian Institute of Science Education and Research Mohali, Sector 81, SAS Nagar, Manauli PO 140306, India\\
$^{4}$School of Earth and Space Exploration, Arizona State University, Tempe, AZ 85287, USA\\
$^{5}$ Laboratory for Atmospheric and Space Physics, University of Colorado, 600 UCB, Boulder, CO 80309, USA \\
$^{6}$ Max Planck Institut für Extraterrestrische Physik, D-85748 Garching, Germany
}

\date{Accepted XXX. Received YYY; in original form ZZZ}

\pubyear{2020}

\begin{document}
\label{firstpage}
\pagerange{\pageref{firstpage}--\pageref{lastpage}}
\maketitle

\begin{abstract}
Our nearest stellar neighbour, Proxima Centauri, is a low mass star with spectral type dM5.5 and hosting an Earth-like planet orbiting within its habitable zone. However, the habitability of the planet depends on the high-energy radiation  of the chromospheric and coronal activity of the host star. 
 We report the \emph{AstroSat}, \emph{Chandra} and \emph{HST} observation of Proxima Centauri carried out as part of multi-wavelength simultaneous observational campaign.  
   Using the soft X-ray data, we probe the different activity states of the star. We investigate the coronal temperatures, emission measures and abundance. Finally, we compare our results with earlier observations of Proxima Centauri.
\end{abstract}

\begin{keywords}
stars: activity  -- stars: coronae --
             stars: low-mass, late-type -- stars: individual: Proxima Centauri
\end{keywords}



\section{Introduction}\label{sec:intro}

The habitability of an extrasolar planet depends not only on its distance from its host but also on the
host's magnetic activity.  The host's high-energy emission in the X-ray and UV-range determines 
the amount of mass loss from the planetary atmosphere \citep{lammer_2003}; given some incident high-energy
flux, evaporation takes place in the form of ``hydrodynamic escape''.   The nearest extrasolar planet
orbits around the star Proxima Centauri, which itself is located in a
 hierarchical triple system consisting of Proxima Centauri and a closer binary $\alpha$ Centauri AB (with
components of spectral type G2V and K1V). Proxima Centauri is an M-dwarf of spectral type dM5.5e \citep{shapley1951}. 

Some solar flares are known to be associated with coronal
mass ejection when suddenly energetic particles are released into interplanetary space. If and when
such particles enter the Earth's magnetosphere, geomagnetic disturbances of various strengths 
are observed.  By analogy, we may expect analogous events in the Proxima Centauri planetary system, and 
recently a superflare event on Proxima Centauri has been reported, 
which would have been barely visible with the naked eye \citep{howard2018}; we summarize some
of the relevant stellar parameters of Proxima Centauri in Tab.~\ref{tab:obs}.

Proxima Centauri, like all flare stars, is a copious producer of X-ray and UV emission. It has, therefore, been observed frequently by several X-ray (and UV satellites) such as \emph{Einstein}, EXOSAT, ROSAT, ASCA, \emph{Chandra} and \emph{XMM-Newton} \citep{haisch_1980, haisch_1983, haisch_1995, gudel_2004}. 
Proxima Centauri's quiescent X-ray emission varies from 4 to 16~$\times$~10$^{26}$~erg/s \citep{haisch_1990}. 
Although its surface area is much smaller than that of our Sun, its 
X-ray emission is usually quite a bit larger. 
Frequent flaring is observed from Proxima Centauri in optical, UV, X-ray and radio wavelengths, with several flares studied in detail over the
over last four decades \citep{haisch_1983, lim_1996, gudel_2002, fuhrmeister_2011, macgregor_2018}.  
Furthermore, based on 15~yr All Sky Automated Survey (ASAS) V-band data \cite{wargelin_2017} found that Proxima Centauri 
shows a $\sim$7 years modulation with similar indications of activity cycle in X-ray wavelength. 

Currently, we do not have a reliable method to estimate the energetic particle output of magnetically
active late-type stars. It is, however, possible to measure the radiative output and assess the planetary evaporation due to incident high-energy radiation.
To assess the full high-energy radiation field between 5 and 3200~\AA \, one has to combine the 
UV and X-ray observations.  Since most of the radiation in the EUV range 
(150 \AA \, $<$ $\lambda$ $<$ 911 \AA ) is obscured by the interstellar medium even for the 
very nearest stars like Proxima Centauri,  we require simultaneous X-ray and UV/optical observations 
as well as emission measure modelling to
reconstruct the coronal contribution to the EUV flux.  
We present a detailed analysis of coronal emission of Proxima Centauri during both quiescent and flaring states observed simultaneously with \emph{AstroSat} Soft X-ray Telescope (SXT; \citealt{singh_2014, singh_2016}), \emph{Chandra} 
Low Energy Transmission Grating (LETG/HRC-S) \citep{brinkman_1987, brinkman_1997} and  \emph{Hubble Space Telescope (HST)} . 

Our paper is structured as follows: in the following section, we describe our observations obtained with \emph{AstroSat} and the data analysis (Section \ref{sec:observation}). In Section \ref{sec:results}, we compare the timing behaviour of Proxima Centauri observed with \emph{AstroSat} and 0th order LETG/HRC-S observations. 
We also briefly present the HST observation of the flare event that was seen simultaneously with \emph{AstroSat} and \emph{Chandra}.
We further discuss the coronal properties of Proxima Centauri and verify the consistency of the results with the LETG/HRC-S observations. In Sections \ref{sec:sum}, we summarise  our findings.

\begin{table}
\centering
\caption{Properties of star systems and the AstroSat Observations.}
\label{tab:obs}
\begin{tabular}{lll} 
\hline\hline
  \textbf{Proxima Centauri}  &\\ [0.5ex] 
\hline
Distance (pc)& 1.3 & a\\
Spectral type& dM5.5e &b\\
Rotation period (d) & 83& c\\

L$_X$ (erg/s) &4-16 $\times$ 10$^{26}$ & d\\
L$_\mathrm{bol}$ (erg/s)& 6 $\times$ 10$^{30}$ & e\\
$\frac{L_X}{L_\mathrm{bol}}$&-4&\\
\hline
\textbf{Proxima Centauri b}  &\\ [0.5ex] 
\hline
Mass (M$_{\oplus}$)&$\sim$1.3&e\\
Orbital period (d) &$\sim$11.2&e\\
Semi-major axis a (AU) & $\sim$0.05&e\\
\hline
\textbf{\emph{AstroSat}}  &\\
\hline
Observation start &31/05/2017 06:08:50\\
Observation end & 01/06/2017 13:51:33\\
Duration (ks)& $\sim$42&\\
Primary instrument & SXT&\\
ObsID& A$03-046$T$01-9000001260$&\\
\hline
\textbf{\emph{Chandra}}  &\\
\hline
Observation start &31/05/2017 16:25:42\\
Observation end & 01/06/2017 05:28:58\\
Duration (ks)&45&\\
Primary instrument & LETG/HRC-S&\\
ObsID& 19708&\\
\hline

\end{tabular}

\footnotesize{\textbf{References:} (a) \cite{van_2007}(b) \cite{boyajian_2012}
(c) \cite{kiraga_2007} (d) \cite{haisch_1990} (d) \cite{reid_2001} (e) \cite{anglada_2016}}

\end{table}

\section{Observations and Data Analysis}
\label{sec:observation}

The data presented in this paper are simultaneous observations of Proxima Centauri carried out with 
India's first space observatory \emph{AstroSat}, the LETG/HRC-S onboard {\it Chandra} and \emph{HST} (ObsID: 14860) between 
31 May - 01 June 2017. The observations cover 19 consecutive \emph{AstroSat} orbits. 
The detailed observation log are provided in Table~\ref{tab:obs}.

\emph{AstroSat} is a multi-wavelength astronomy mission, carrying five 
multi-band payloads \citep{agrawal_2006, singh_2014, agrawal_2017}. 
The Soft X-ray Telescope aboard \emph{AstroSat} covers the 0.3-7.0 keV energy band with an 
effective area of $\sim$90 cm$^2$ at 1.5 keV \citep{singh_2017a}.  
The SXT is a grazing incidence soft X-ray telescope with a focal length of 200~cm, equipped with a thermo-electrically cooled CCD detector in its focal plane \citep{singh_2014, singh_2016, singh_2017a,singh_2017b}. The SXT level 1 data from individual orbits are received at the SXT Payload Operation Centre (POC) 
and are processed using the SXTPIPELINE version 1.4b at the POC. Using the SXTPIPELINE the events are extracted 
and time tagged, and the raw coordinates are transformed to sky coordinates. The pipeline also carries out bias subtraction, bad pixel flagging, event grading and pulse height amplitude 
reconstruction for each event. So-called ``Good Time Interval''  files for each orbit are selected and merged 
event files are created\footnote{\href{http://www.tifr.res.in/~astrosat_sxt/dataanalysis.html}{http://www.tifr.res.in/$\sim$astrosat$\_s$xt/dataanalysis.html}}. 

\begin{figure}
\centering
\includegraphics[width=0.475\textwidth]{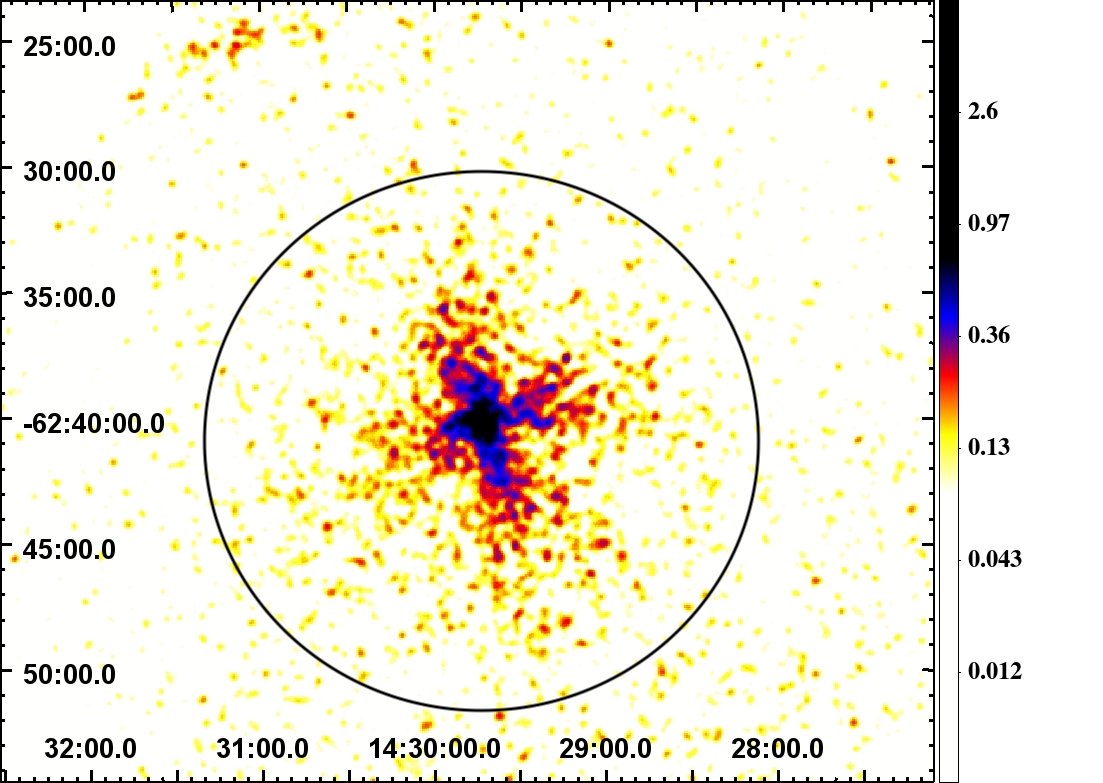}
\caption{Soft X-ray image of Proxima Centauri merged soft X-ray data in the 0.5-4 keV energy band. 
The black circle represents the source signal extracted from a circular region with 12' radius.}
\label{fig:ds9}
\end{figure}

We used XSELECT version 2.4c in HEASOFT 6.24 to extract the light curve for the regions
centered on the sources. 
The SXT light curves and spectra were obtained using standard filtering 
criteria\footnote{\href{http://astrosat-ssc.iucaa.in/?q=data\_and\_analysisl}{http://astrosat-ssc.iucaa.in/?q=data\_and\_analysis}}. We used a blank-sky SXT spectrum provided by the SXT team for modeling the
background spectrum. 
We used the XSPEC-compatible RMF and ARF files 
for SXT provided by the \emph{AstroSat} Science Support cell$^3$. 
These were sxt$\_$pc$\_$excl00$\_$v04$\_$a.arf and sxt$\_$pc$\_$mat$\_$g0to12.rmf. 
The spectra were binned to a minimum of 25 counts per energy bin.
The spectral analysis was carried out with XSPEC version 12.9.0  \citep{xspec} after dividing the data into
periods of quiescence and flaring (see Sections \ref{ssec:lc} and \ref{ssec:spec}).

\begin{figure*}
\centering
\includegraphics[width=\textwidth]{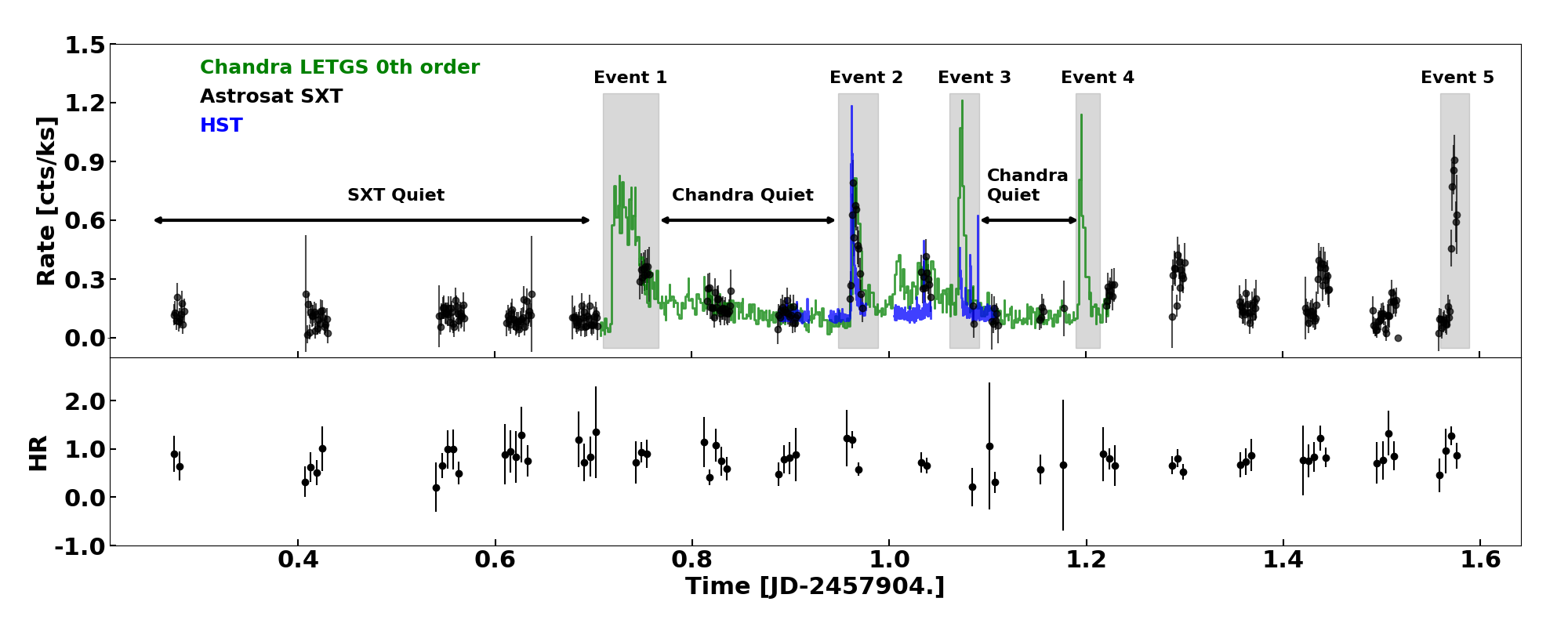}
\caption{\label{fig:lc} Top panel: \emph{Astrosat} SXT (black) and \emph{Chandra} LETG/HRC-S 0th order (in green) light curves of Proxima Centauri plotted in 0.3-3.0 keV energy band binned to 100~s. The scaled \emph{HST} light curve is represented in blue. 
Plotted in the bottom panel is the \emph{Astrosat} SXT hardness ratio binned to 500~s.  The time segments 
corresponding to the flare-like events are represented by the vertical blocks.}
\label{lightcurve}
\end{figure*}

For the {\it Chandra} data, we followed the standard CIAO analysis guide\footnote{\href{http://cxc.harvard.edu/ciao/guides/}{http://cxc.harvard.edu/ciao/guides/}} for reducing {\it Chandra} LETG/HRC-S data. 
We used CIAO version 4.10 of the data processing software. According to the standard data processing procedure the Good Time Interval files were created. 

Proxima Centauri  was simultaneously observed using the STIS spectrograph aboard \textit{HST} with the E140M echelle grating. To produce flux and wavelength calibrated spectra, we used the standard \texttt{calSTIS} pipeline version FIXME. We split the HST observations into sub-exposures with the \texttt{calSTIS inttag} function at a cadence of FIXME, resulting in individual spectra for each of these sub-exposures. From these spectra, we integrated flux from 1200-1700 \AA\ region to generate a broadband FUV light curve.

\section{Results}\label{sec:results}

To provide an impression of what \emph{AstroSat} data look like, we show our \emph{AstroSat}-SXT 
image of Proxima Centauri in Fig.~\ref{fig:ds9}.
The source signal was extracted from a circular region with 12$\arcmin$ radius centered on the proper-motion corrected and nominal position of Proxima Centauri.

\subsection{X-ray light curves}\label{ssec:lc}
\label{sec:lc}

In Fig.~\ref{fig:lc} (black data points), we plot the background-subtracted SXT light curves for 
Proxima Centauri binned by 100~s over the
entire 110~ks observing period by \emph{AstroSat}; the obtained zeroth-order LETG/HRC-S
light curve is also shown as green data points.  Fig.~\ref{fig:lc} 
clearly shows the different observation patterns followed by  \emph{AstroSat} and {\it Chandra};
the \emph{AstroSat} data stream is frequently interrupted by Earth blocks, while the
{\it Chandra} data stream is continuous.  Fortuitously, the count rates measured by
both instruments are very similar. 
Although the HRC-S has no
useful energy resolution in 0th order, we can study spectral
variability by using the SXT data to create hardness ratios (HR). It is
defined as the ratio between the counts in the hard band (1.0-3.0~keV) 
and the soft band (0.3-1.0~keV) we can study spectral variability. In Fig.~\ref{fig:lc} bottom panel, the HR binned to 500~s is plotted. 
Due to the sparse sampling of the HR plots, we do not see any heating related trends.

Several energetic events can be seen in both \emph{AstroSat} and 
{\it Chandra} light curves; we call them Events 1-5. 
There are four flare-like events seen in the {\it Chandra} light curve we call them Events 1-4. 
The peak count rate of Event 1, 3 and 4 observed in {\it Chandra} light curves are $\sim$0.75, $\sim$1.23 and $\sim$1.18 cts s$^{-1}$, respectively. 
Event~2 which
occurred around JD$\sim$2457904.94  was covered by both
X-ray instruments as well as \emph{HST}. During Event 2
the count rate increased from the quiescent value of $\sim0.13$ cts s$^{-1}$ 
to $\sim0.80$ cts s$^{-1}$ in both {\it Chandra} and SXT lightcurves in 0.5-4.0 keV range. 
Correspondingly, during the flare the \emph{HST} lightcurve shows that the normalised flux increases by two orders of magnitude. The Event~2 occurs over $\sim$2700s producing integrated energy of 
$\sim$4.7$\times10^{30}$ erg in 0.5-4.0 keV (see Section \ref{ssec:spec} for discussion of spectral fitting). 
The energetics of the Event 2 indicates a moderate flare in comparison to the 
very strong flare studied by \cite{gudel_2004} with a total energy of $\approx 2 \times 10^{32}$ erg and several other 
flares with energies of the order of $10^{31}$ erg studied by \cite{fuhrmeister_2011}. 
Furthermore, we note an increased activity state of Proxima Centauri
towards the end of \emph{AstroSat} observations which we call Event 5. The count rate reaches a maximum of 
$\approx$ 0.90 cts s$^{-1}$.  However, since the observations ended during the event we did not carry out further analysis.

\begin{figure*}
\centering
\includegraphics[width=0.49\textwidth]{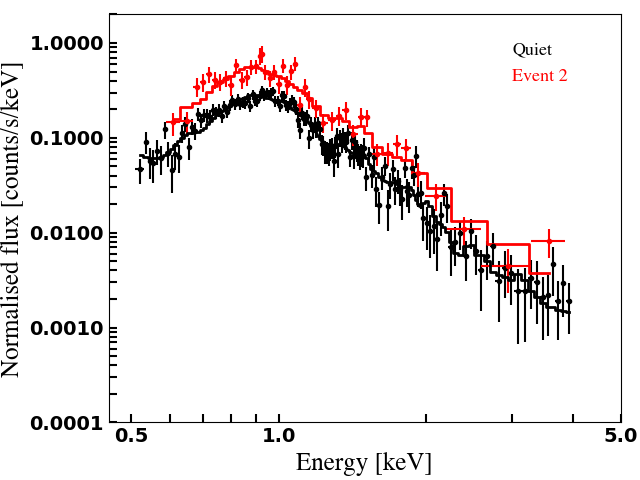}
\includegraphics[width=0.49\textwidth]{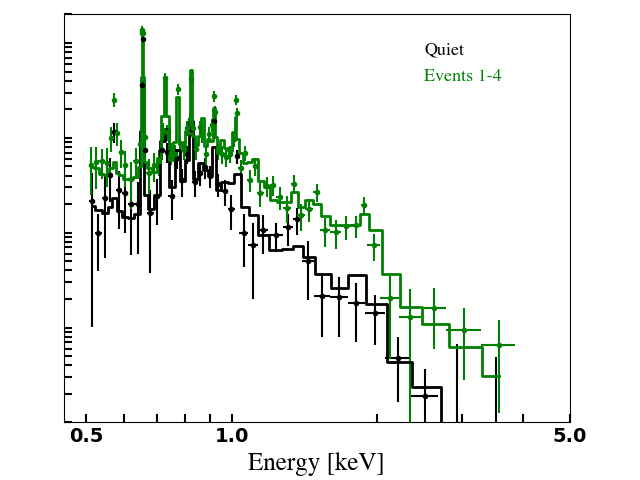}

\caption{\label{fig:spec} Left panel: {\it AstroSat} SXT spectra of Proxima Centauri during quiescent (black) and flare (red) in 0.5-4.0 keV. Right panel: {\it Chandra} LETG/HRC-S spectra integrated spectra of 
quiescent duration (black) and combined flaring events 1-4 (green).}
\label{lightcurve}
\end{figure*}

\begin{table*}
\centering
\caption{Three-temperature fit to the {\it AstroSat} SXT and {\it Chandra} LETG/HRC-S spectra in the energy range 0.5-4.0 keV during different activity states allowing individual elemental abundance and emission measure to vary. Errors given are 1$\sigma$ errors.}
\label{tab:fit}
\begin{tabular}{llllllll} 
    \hline
 \multicolumn{1}{l}{Parameter}  &
      \multicolumn{2}{c}{LETG/HRC-S} &
      \multicolumn{2}{c}{\emph{AstroSat}}   \\
      \hline
      
    &Quiet& Events 1-4& Quiet & Event 2\\
    \hline
T$_1$ & 0.5& 0.5&0.5&0.5\\
$[$keV$]$ & && &\\
T$_2$ & 1.0&1.0&1.0&1.0\\
$[$keV$]$ &  &&&\\
T$_3$ & 2.0&2.0&2.0&2.0\\
$[$keV$]$ &  &&&\\
EM$_1$ & 2.16 $^{+1.71}_{-1.61}$&  2.04$^{+1.38}_{-0.97}$&3.99$^{+2.28}_{-1.62}$&9.88$^{+7.47}_{-4.43}$\\
$[10^{49}$ cm$^{-2}]$ &&&\\
EM$_2$ & 2.64$^{+1.15}_{-1.42}$&  3.76$^{+4.92}_{-3.58}$&13.21$^{+4.93}_{-5.41}$&15.47$^{+6.63}_{-8.43}$\\
$[10^{49}$ cm$^{-2}]$ &&&\\
EM$_3$ & 5.93$^{+1.40}_{-1.28}$&   5.14$^{+2.79}_{-3.93}$&10.81$^{+2.42}_{-3.64}$&12.32$^{+4.32}_{-6.99}$\\
$[10^{49}$ cm$^{-2}]$ &&&\\
\hline
Abundance$^{\dagger}$ &  0.23$^{+0.86}_{-0.12} $ & 0.35$^{+0.08}_{-0.09}$ &0.38$^{+0.14}_{-0.10}$&0.32$^{+0.31}_{-0.16}$\\
\hline
Flux $\times 10^{-12}$&  3.52$\pm$0.58 & 7.85$\pm$ 0.75  &5.13 $\pm$0.72& 15.59 $\pm$1.56\\
$[\textrm{erg}~\textrm{s}^{-1}~\textrm{cm}^{-2}]$ &&&&\\
L$_X$& 0.71$\pm$0.11 &      1.58$\pm$0.15    &1.03$\pm$0.15&3.15$\pm$0.32 \\
 $[10^{27} \textrm{erg}~\textrm{s}^{-1}]$ &&&&\\
 $\chi^2$&1.77&1.93&1.12&1.23\\
\hline
\end{tabular}

\footnotesize{$^{\dagger}$Relative to solar photosphere}

\end{table*}

\subsection{Spectra}\label{ssec:spec}

We extracted the quiescent and flaring interval spectra of Proxima Centauri 
to study the emission measure and coronal elemental abundance changes as a result of the flares. In Fig.~\ref{fig:spec} (left panel), we plot the SXT spectra during quiescence (black data points) and  event 2 (red data points). 

The {\it Chandra} LETG/HRC-S zeroth-order light curve shows a large number of flaring events. 
The event 2 observed by \emph{AstroSat} SXT has been observed by {\it Chandra} 
LETG/HRC-S as well, however, the signal-to-noise ratio for LETG/HRC-S spectra was not adequate 
for the spectral fit. 
Hence we combined all the LETG flare data (events 1-4) 
for fitting and selected two time periods in the remaining data (marked 
as "Chandra Quiet" in Figure \ref{fig:lc}) to create a quasi-quiescent spectrum.
In Figure~\ref{fig:spec} right panel, we show the {\it Chandra} integrated spectra of the quiescent and combined events 1-4.
We fit the LETG/HRC-S spectra in 0.206-7.0 keV range for the positive order and 0.252-7.0 keV range for the negative order independently for the 
flare and quiescent time intervals.

Using XSPEC, we performed a global fit to the SXT spectra using APEC plasma models
\footnote{\href{https://heasarc.gsfc.nasa.gov/xanadu/xspec/manual/node135.html}{https://heasarc.gsfc.nasa.gov/xanadu/xspec/manual/node135.html}} to the 
SXT spectra. To compare the quiescent and the flaring state we defined a fixed temperature grid with values 0.5, 1.0 and 2.0 keV (5.7, 11.5 and 23~MK, respectively). 
In these fits, the determined abundances are relative to solar 
values \cite{grevesse_1998}, and all three
temperature components share the same abundance. 
We allowed the abundance to vary independently to fit the quiescent and flare spectra. 
Assuming a distance of $\sim$1.3 pc, we also calculate the emission measure from the 
component's normalisation. Similar to SXT spectra, we fit the quiescent and the events 1-4 observed with {\it Chandra} 
with 3 temperature APEC models allowing the abundance to vary freely. 
The resulting fit parameters are given in Table ~\ref{tab:fit}.

Although the 3 temperature component APEC may be adequate to interpret the spectra observed with SXT, 
the assumption of three discrete temperatures is not realistic to physically describe
the stellar coronae. Hence we also experimented with a more complex model such as  a 
c6pmekl \footnote{\href{https://heasarc.gsfc.nasa.gov/xanadu/xspec/manual/node149.html}{https://heasarc.gsfc.nasa.gov/xanadu/xspec/manual/node149.html}} which uses 6th order Chebyshev polynomial describing 
differential emission measure \citep{lemen_1989, singh_1996}. However,  we found that the 3 temperature model described above provides a reasonable description of our data.

During a flare, 
fresh chromospheric material unaffected by elemental differentiation ( First ionisation potential effect also known as "FIP effect") is expected to be evaporated and transported into 
the corona, thus changing the coronal emission measure and potentially some 
element abundances  \citep{ottmann_1996, benz_2010}.  
In a moderately active M star such as Proxima Centauri, during quiescent 
emission, a pronounced anti-FIP effect is expected  \citep{Wood_2010, fuhrmeister_2011}, but uncertainties for the 
abundances during quiescence and flaring were too large in our fits 
to detect any differences.

\begin{figure}
\centering
\includegraphics[width=0.495\textwidth]{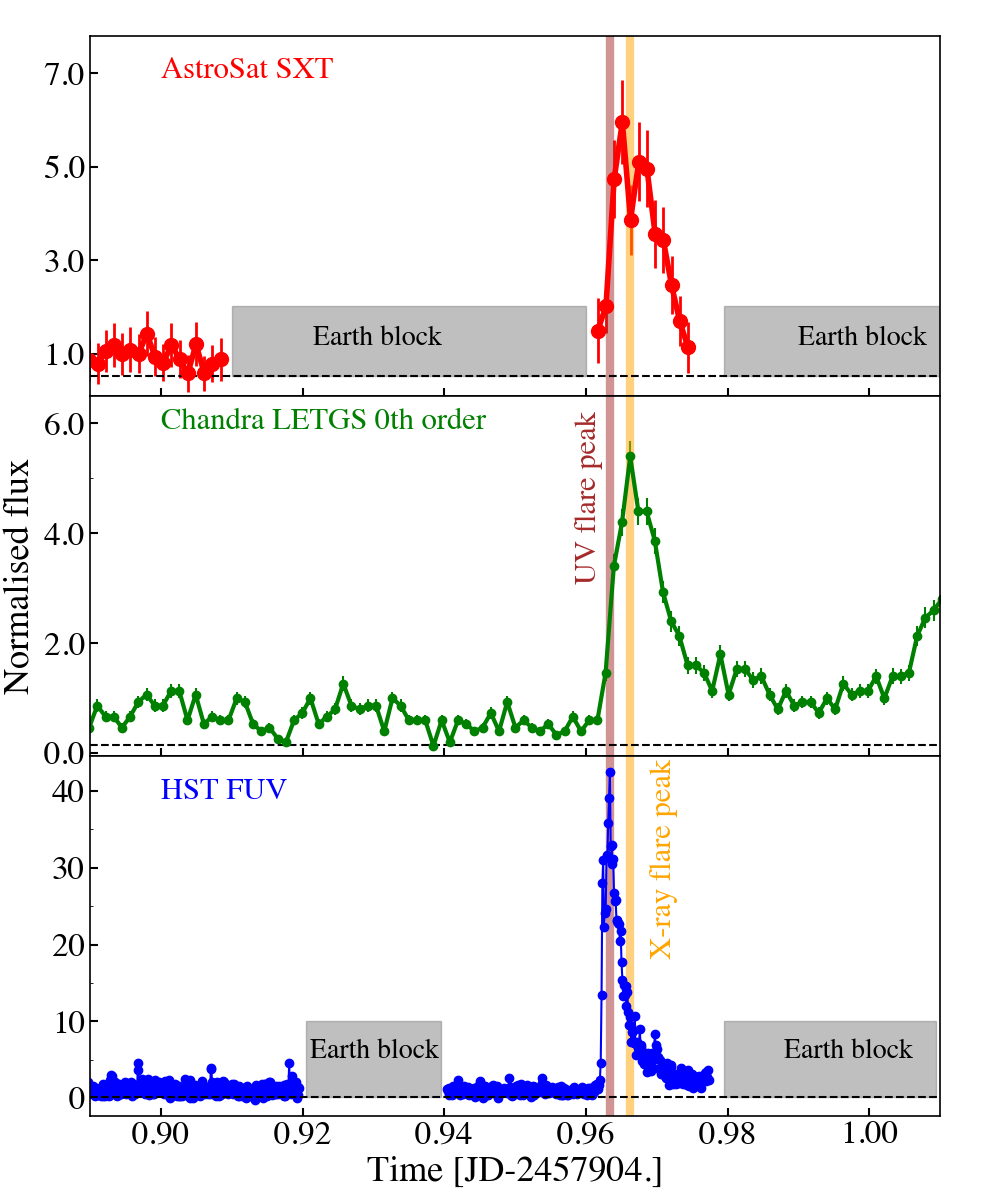}
\caption{\label{fig:neup} Event 2 simultaneously observed  with the \emph{AstroSat} SXT in 0.5-4.0 keV (top panel), {\it Chandra} LETG/HRC-S zeroth-order in 0.1-12.0 keV (middle panel) and HST FUV in 120- 170nm (bottom panel). }
\label{lightcurve}
\end{figure}

Since the quantum efficiency and the effective area of X-ray instruments are not uniform over energy bands, 
we compare the {\it Chandra} LETG/HRC-S and {\it AstroSat} SXT in smaller segments of energy bandpasses. 
We used the quiescent spectra of both the instruments in five energy bandpasses 0.5-1.0 keV, 1.0-1.5 keV, 1.5-2.0 keV, 2.0-4.0 keV,  and 4.0-7.0 keV. 
We fit 3 temperature APEC models and ignored all the data points out of the energy band of interest. 
We note that the LETG/HRC-S above 2 keV is not useful due to statistical limitations. 
The  flux ratio between the \emph{AstroSat} SXT and \emph{Chandra} LETG/HRC-S produces a difference 
$>$1.5$\sigma$. The difference in the fluxes could be inherent due to the difference in the time periods chosen for the spectra. 
Furthermore, since the calibration to convert the counts into absolute fluxes are not very accurate, 
this can also result in the discrepancies when we compare the fluxes from different instruments.

\subsection{Simultaneous observation}

In Figure~\ref{fig:neup}, we show a close-up view of Event~1 observed simultaneously with \emph{AstroSat} SXT, {\it Chandra} LETG/HRC-S and \emph{HST} binned to 100s each. 
The vertical lines in brown and yellow represents the UV and X-ray flare peaks, respectively. 
The amplitudes of the \emph{HST} light curves  are scaled to be consistent with X-ray light curves. 
A detailed analysis of the \emph{HST} data will be presented in Schneider et al (in preparation). 
However, a visual inspection of Fig.~\ref{fig:neup} indicates that the \emph{HST} FUV light curve 
precedes the X-ray light curves by about a few minutes. 
This may indicate that the presence of Neupert effect \citep{neupert_1968} during the rising phase of Event~2. 
Using the Z-transformed Discrete Correlation Function (ZDCF) \citep{alexander_1997},  
the cross-correlation of the \emph{HST} and the X-ray light curves produces a 
time-lag between the peaks to be 300-400 s, indicating that due to electron impingement the UV light curve precedes 
the thermal radiation.

\subsection{Comparison}\label{sec:compare}
\begin{figure}
\centering
\includegraphics[width=0.48\textwidth]{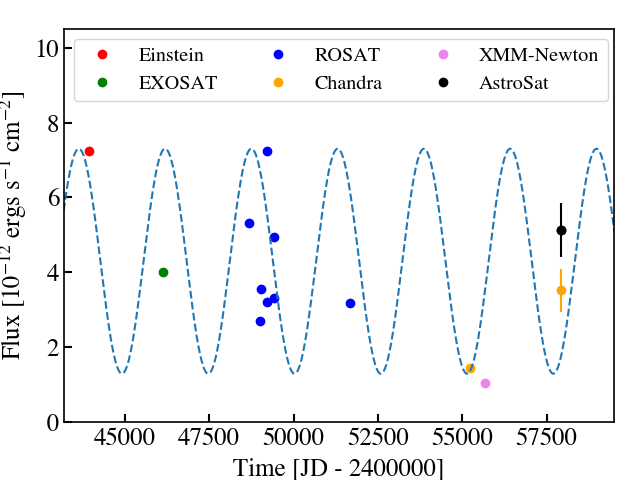}
\caption{\label{fig:comp} Quiescent emission flux as a function of time in 0.3-4.0 keV energy range observed by several X-ray missions between 1979-2017. An arbitrary sine curve with  a period of 
7 yr is represented as the blue dashed curve.}
\label{lightcurve}
\end{figure}

The quiescent emission measured with \emph{Einstein}, EXOSAT and ROSAT in 0.3-4.0 keV 
energy ranges are 7.2$\times 10^{-12}$, 4-16$\times 10^{-12}$ and 2.6-7.2$\times 10^{-12}$ erg s$^{-1}$ cm$^{-2}$, respectively.  
The quiescent X-ray flux observed by \emph{Einstein} and EXOSAT was reported by \cite{haisch_1980} and \cite{haisch_1990}, respectively.
We used the ROSAT observations listed in the ROSAT Positive Sensitive Proportional Counter source catalogue \cite{voges_1996, voges_2000}. 
Assuming a coronal temperature of 1~keV, sub-solar abundance of 0.5 and using WebPIMMS, we computed energy conversion factor 
ECF$_{PSPC}$ = 6.4$\times$ $10^{-12}$ erg cm$^{-2}$ counts$^{-1}$ to convert the observed ROSAT 
counts to flux in canonical 0.1-2.4 keV ROSAT energy band.
Furthermore, we use the X-ray flux reported by \cite{fuhrmeister_2011} based on \emph{XMM-Newton} observations.
In Figure~\ref{fig:comp}, we plot quiescent flux measured between 1979 and 2017 as a 
function of time. Furthermore, we assume similar temperature and abundance as before and used WebPIMMS to estimate the flux in the 0.3-4.0 keV energy range. 
The flux obtained from \emph{AstroSat} is consistent with the previous estimates of quiescent flux for Proxima Centauri.

While a clear cyclic behaviour with a period of $\sim$7.0 years is observed  for Proxima Centauri in ASAS observations, a similar variation in the X-ray data is not apparent immediately. 
However, \cite{wargelin_2017} reported the evidence for a 7-yr stellar cycle using observations from several X-ray missions, principally \emph{Swift}. The temporal coverage of X-ray observation, the energy resolution and uncertainties due cross-calibration restrict a detailed analysis of the cyclic behaviour of Proxima Centauri. However, we performed a periodogram analysis of all the compiled X-ray data, but  we were unable to obtain a  significant peak indicating a period.  
We used the 7-yr period obtained from the optical light curve obtained based on ASAS observations reported by \citep{suarez_2016}.  We assumed that the optical period also 
applies to the X-ray data and plot a sinusoidal curve with a period of 7-yr in Fig.~\ref{fig:comp}.  Furthermore, assuming that the optical cycle also applies to the X-ray cycle, we carried out a correlation analysis between the two datasets. We used the fitted optical light curve at each X-ray observation, to relate the X-ray observations to the measured optical magnitude. We computed the linear Pearson correlation coefficient ($\rho$) of -0.42 between the X-ray flux and the optical magnitude with a two-tailed probability value of 0.02$\%$. The results suggest an anti-correlation between the X-ray and optical data. However, obtaining additional X-ray data from missions such as \emph{AstroSat} will be valuable to obtain a precise cyclic variation of Proxima Centauri. 

\section{Summary}\label{sec:sum}
To summarise,
\begin{itemize}
    \item We have presented the light curves and spectra of Proxima Centauri obtained from the SXT onboard \emph{AstroSat}.  
    The observation was carried out as a part of the simultaneous observational campaign with \emph{Chandra} and \emph{HST}.
     \item Both SXT and \emph{Chandra} LETG/HRC-S zeroth-order light curves cover a flare-like event. 
    \item The timing behaviour of the flare-like event shows Neupert effect similar to the solar flares 
    where the UV light curve precedes the soft X-ray light curves.
    \item Three-temperature component models provide the best fit to SXT spectra with sub-solar abundance. 
    \item The emission measure distribution model peaks around 1 keV. 
    \item The \emph{AstroSat} quiescent X-ray flux in 0.3-4.0 keV energy range is consistent with previous estimates of quiescent flux.

\end{itemize}

\section*{Acknowledgements}

This publication uses the data from the AstroSat mission of the Indian Space Research Organisation (ISRO). This work has used the data from the Soft X-ray Telescope (SXT) 
The SXT POCs at TIFR are thanked for verifying and releasing the data via the ISSDC data archive and providing the necessary software tools. 
This research has made use of the data and software obtained from NASA's High Energy Astrophysics Science Archive Research Center (HEASARC), a service of Goddard Space Flight Center and the Smithsonian Astrophysical Observatory. The use of the XRT Data Analysis Software (XRTDAS) developed under the responsibility of the ASI Science Data Center (ASDC), is gratefully acknowledged. 
This work has made use of observations with {\it Chandra} and \emph{HST}. This work was also partially supported by a Leverhulme Trust Research Project Grant.


\bibliographystyle{mnras}
\bibliography{paper} 

\begin{thebibliography}{}
\makeatletter
\relax
\def\mn@urlcharsother{\let\do\@makeother \do\$\do\&\do\#\do\^\do\_\do\%\do\~}
\def\mn@doi{\begingroup\mn@urlcharsother \@ifnextchar [ {\mn@doi@}
  {\mn@doi@[]}}
\def\mn@doi@[#1]#2{\def\@tempa{#1}\ifx\@tempa\@empty \href
  {http://dx.doi.org/#2} {doi:#2}\else \href {http://dx.doi.org/#2} {#1}\fi
  \endgroup}
\def\mn@eprint#1#2{\mn@eprint@#1:#2::\@nil}
\def\mn@eprint@arXiv#1{\href {http://arxiv.org/abs/#1} {{\tt arXiv:#1}}}
\def\mn@eprint@dblp#1{\href {http://dblp.uni-trier.de/rec/bibtex/#1.xml}
  {dblp:#1}}
\def\mn@eprint@#1:#2:#3:#4\@nil{\def\@tempa {#1}\def\@tempb {#2}\def\@tempc
  {#3}\ifx \@tempc \@empty \let \@tempc \@tempb \let \@tempb \@tempa \fi \ifx
  \@tempb \@empty \def\@tempb {arXiv}\fi \@ifundefined
  {mn@eprint@\@tempb}{\@tempb:\@tempc}{\expandafter \expandafter \csname
  mn@eprint@\@tempb\endcsname \expandafter{\@tempc}}}

\bibitem[\protect\citeauthoryear{{Agrawal}}{{Agrawal}}{2006}]{agrawal_2006}
{Agrawal} P.~C.,  2006, \mn@doi [Advances in Space Research]
  {10.1016/j.asr.2006.03.038}, \href
  {http://adsabs.harvard.edu/abs/2006AdSpR..38.2989A} {38, 2989}

\bibitem[\protect\citeauthoryear{{Agrawal}}{{Agrawal}}{2017}]{agrawal_2017}
{Agrawal} P.~C.,  2017, \mn@doi [Journal of Astrophysics and Astronomy]
  {10.1007/s12036-017-9449-6}, \href
  {http://adsabs.harvard.edu/abs/2017JApA...38...27A} {38, 27}

\bibitem[\protect\citeauthoryear{{Alexander}}{{Alexander}}{1997}]{alexander_1997}
{Alexander} T.,  1997, in {Maoz} D.,  {Sternberg} A.,   {Leibowitz} E.~M.,
  eds,  Astrophysics and Space Science Library Vol. 218, Astronomical Time
  Series. p.~163, \mn@doi{10.1007/978-94-015-8941-3_14}

\bibitem[\protect\citeauthoryear{{Anglada-Escud{\'e}}
  et~al.,}{{Anglada-Escud{\'e}} et~al.}{2016}]{anglada_2016}
{Anglada-Escud{\'e}} G.,  et~al., 2016, \mn@doi [\nat] {10.1038/nature19106},
  \href {http://adsabs.harvard.edu/abs/2016Natur.536..437A} {536, 437}

\bibitem[\protect\citeauthoryear{{Arnaud}}{{Arnaud}}{1996}]{xspec}
{Arnaud} K.~A.,  1996, in {Jacoby} G.~H.,  {Barnes} J.,  eds, ASP Conf. Ser.
  101: Astronomical Data Analysis Software and Systems V. pp 17--+

\bibitem[\protect\citeauthoryear{{Benz} \& {G{\"u}del}}{{Benz} \&
  {G{\"u}del}}{2010}]{benz_2010}
{Benz} A.~O.,  {G{\"u}del} M.,  2010, \mn@doi [\araa]
  {10.1146/annurev-astro-082708-101757}, \href
  {http://adsabs.harvard.edu/abs/2010ARA%26A..48..241B} {48, 241}

\bibitem[\protect\citeauthoryear{{Boyajian} et~al.,}{{Boyajian}
  et~al.}{2012}]{boyajian_2012}
{Boyajian} T.~S.,  et~al., 2012, \mn@doi [\apj] {10.1088/0004-637X/757/2/112},
  \href {http://adsabs.harvard.edu/abs/2012ApJ...757..112B} {757, 112}

\bibitem[\protect\citeauthoryear{{Brinkman}, {van Rooijen}, {Bleeker},
  {Dijkstra}, {Heise}, {de Korte}, {Mewe}  \& {Paerels}}{{Brinkman}
  et~al.}{1987}]{brinkman_1987}
{Brinkman} A.~C.,  {van Rooijen} J.~J.,  {Bleeker} J.~A.~M.,  {Dijkstra} J.~H.,
   {Heise} J.,  {de Korte} P.~A.~J.,  {Mewe} R.,   {Paerels} F.,  1987,
  Astrophysical Letters and Communications, \href
  {https://ui.adsabs.harvard.edu/abs/1987ApL%26C..26...73B} {26, 73}

\bibitem[\protect\citeauthoryear{{Brinkman} et~al.,}{{Brinkman}
  et~al.}{1997}]{brinkman_1997}
{Brinkman} A.~C.,  et~al., 1997, in {Hoover} R.~B.,  {Walker} A.~B.,  eds,
  \procspie Vol. 3113, Grazing Incidence and Multilayer X-Ray Optical Systems.
  pp 181--192, \mn@doi{10.1117/12.278846}

\bibitem[\protect\citeauthoryear{{Fuhrmeister}, {Lalitha}, {Poppenhaeger},
  {Rudolf}, {Liefke}, {Reiners}, {Schmitt}  \& {Ness}}{{Fuhrmeister}
  et~al.}{2011}]{fuhrmeister_2011}
{Fuhrmeister} B.,  {Lalitha} S.,  {Poppenhaeger} K.,  {Rudolf} N.,  {Liefke}
  C.,  {Reiners} A.,  {Schmitt} J.~H.~M.~M.,   {Ness} J.-U.,  2011, \mn@doi
  [\aap] {10.1051/0004-6361/201117447}, \href
  {http://adsabs.harvard.edu/abs/2011A%26A...534A.133F} {534, A133}

\bibitem[\protect\citeauthoryear{{Grevesse} \& {Sauval}}{{Grevesse} \&
  {Sauval}}{1998}]{grevesse_1998}
{Grevesse} N.,  {Sauval} A.~J.,  1998, \ssr, 85, 161

\bibitem[\protect\citeauthoryear{{G{\"u}del}, {Audard}, {Skinner}  \&
  {Horvath}}{{G{\"u}del} et~al.}{2002}]{gudel_2002}
{G{\"u}del} M.,  {Audard} M.,  {Skinner} S.~L.,   {Horvath} M.~I.,  2002,
  \mn@doi [\apjl] {10.1086/345404}, \href
  {http://adsabs.harvard.edu/abs/2002ApJ...580L..73G} {580, L73}

\bibitem[\protect\citeauthoryear{{G{\"u}del}, {Audard}, {Reale}, {Skinner}  \&
  {Linsky}}{{G{\"u}del} et~al.}{2004}]{gudel_2004}
{G{\"u}del} M.,  {Audard} M.,  {Reale} F.,  {Skinner} S.~L.,   {Linsky} J.~L.,
  2004, \mn@doi [\aap] {10.1051/0004-6361:20031471}, \href
  {http://adsabs.harvard.edu/abs/2004A%26A...416..713G} {416, 713}

\bibitem[\protect\citeauthoryear{{Haisch} \& {Linsky}}{{Haisch} \&
  {Linsky}}{1980}]{haisch_1980}
{Haisch} B.~M.,  {Linsky} J.~L.,  1980, \mn@doi [\apjl] {10.1086/183193}, \href
  {http://adsabs.harvard.edu/abs/1980ApJ...236L..33H} {236, L33}

\bibitem[\protect\citeauthoryear{{Haisch}, {Linsky}, {Bornmann}, {Stencel},
  {Antiochos}, {Golub}  \& {Vaiana}}{{Haisch} et~al.}{1983}]{haisch_1983}
{Haisch} B.~M.,  {Linsky} J.~L.,  {Bornmann} P.~L.,  {Stencel} R.~E.,
  {Antiochos} S.~K.,  {Golub} L.,   {Vaiana} G.~S.,  1983, \mn@doi [\apj]
  {10.1086/160866}, \href {http://adsabs.harvard.edu/abs/1983ApJ...267..280H}
  {267, 280}

\bibitem[\protect\citeauthoryear{{Haisch}, {Butler}, {Foing}, {Rodono}  \&
  {Giampapa}}{{Haisch} et~al.}{1990}]{haisch_1990}
{Haisch} B.~M.,  {Butler} C.~J.,  {Foing} B.,  {Rodono} M.,   {Giampapa} M.~S.,
   1990, \aap, \href {http://adsabs.harvard.edu/abs/1990A%26A...232..387H}
  {232, 387}

\bibitem[\protect\citeauthoryear{{Haisch}, {Antunes}  \& {Schmitt}}{{Haisch}
  et~al.}{1995}]{haisch_1995}
{Haisch} B.,  {Antunes} A.,   {Schmitt} J.~H.~M.~M.,  1995, \mn@doi [Science]
  {10.1126/science.268.5215.1327}, \href
  {http://adsabs.harvard.edu/abs/1995Sci...268.1327H} {268, 1327}

\bibitem[\protect\citeauthoryear{{Howard} et~al.,}{{Howard}
  et~al.}{2018}]{howard2018}
{Howard} W.~S.,  et~al., 2018, \mn@doi [\apjl] {10.3847/2041-8213/aacaf3},
  \href {http://esoads.eso.org/abs/2018ApJ...860L..30H} {860, L30}

\bibitem[\protect\citeauthoryear{{Kiraga} \& {Stepien}}{{Kiraga} \&
  {Stepien}}{2007}]{kiraga_2007}
{Kiraga} M.,  {Stepien} K.,  2007, \actaa, \href
  {http://adsabs.harvard.edu/abs/2007AcA....57..149K} {57, 149}

\bibitem[\protect\citeauthoryear{{Lammer}, {Selsis}, {Ribas}, {Guinan}, {Bauer}
   \& {Weiss}}{{Lammer} et~al.}{2003}]{lammer_2003}
{Lammer} H.,  {Selsis} F.,  {Ribas} I.,  {Guinan} E.~F.,  {Bauer} S.~J.,
  {Weiss} W.~W.,  2003, \mn@doi [\apjl] {10.1086/380815}, \href
  {http://adsabs.harvard.edu/abs/2003ApJ...598L.121L} {598, L121}

\bibitem[\protect\citeauthoryear{{Lemen}, {Mewe}, {Schrijver}  \&
  {Fludra}}{{Lemen} et~al.}{1989}]{lemen_1989}
{Lemen} J.~R.,  {Mewe} R.,  {Schrijver} C.~J.,   {Fludra} A.,  1989, \mn@doi
  [\apj] {10.1086/167508}, \href
  {http://adsabs.harvard.edu/abs/1989ApJ...341..474L} {341, 474}

\bibitem[\protect\citeauthoryear{{Lim}, {White}  \& {Slee}}{{Lim}
  et~al.}{1996}]{lim_1996}
{Lim} J.,  {White} S.~M.,   {Slee} O.~B.,  1996, \mn@doi [\apj]
  {10.1086/177025}, \href
  {https://ui.adsabs.harvard.edu/abs/1996ApJ...460..976L} {460, 976}

\bibitem[\protect\citeauthoryear{{MacGregor}, {Weinberger}, {Wilner},
  {Kowalski}  \& {Cranmer}}{{MacGregor} et~al.}{2018}]{macgregor_2018}
{MacGregor} M.~A.,  {Weinberger} A.~J.,  {Wilner} D.~J.,  {Kowalski} A.~F.,
  {Cranmer} S.~R.,  2018, \mn@doi [\apjl] {10.3847/2041-8213/aaad6b}, \href
  {https://ui.adsabs.harvard.edu/abs/2018ApJ...855L...2M} {855, L2}

\bibitem[\protect\citeauthoryear{{Neupert}}{{Neupert}}{1968}]{neupert_1968}
{Neupert} W.~M.,  1968, \mn@doi [\apjl] {10.1086/180220}, \href
  {https://ui.adsabs.harvard.edu/abs/1968ApJ...153L..59N} {153, L59}

\bibitem[\protect\citeauthoryear{{Ottmann} \& {Schmitt}}{{Ottmann} \&
  {Schmitt}}{1996}]{ottmann_1996}
{Ottmann} R.,  {Schmitt} J.~H.~M.~M.,  1996, \aap, \href
  {http://adsabs.harvard.edu/abs/1996A%26A...307..813O} {307, 813}

\bibitem[\protect\citeauthoryear{{Reid}, {Burgasser}, {Cruz}, {Kirkpatrick}  \&
  {Gizis}}{{Reid} et~al.}{2001}]{reid_2001}
{Reid} I.~N.,  {Burgasser} A.~J.,  {Cruz} K.~L.,  {Kirkpatrick} J.~D.,
  {Gizis} J.~E.,  2001, \mn@doi [\aj] {10.1086/319418}, \href
  {http://adsabs.harvard.edu/abs/2001AJ....121.1710R} {121, 1710}

\bibitem[\protect\citeauthoryear{{Shapley}}{{Shapley}}{1951}]{shapley1951}
{Shapley} H.,  1951, \mn@doi [Proceedings of the National Academy of Science]
  {10.1073/pnas.37.1.15}, \href {http://esoads.eso.org/abs/1951PNAS...37...15S}
  {37, 15}

\bibitem[\protect\citeauthoryear{{Singh}, {White}  \& {Drake}}{{Singh}
  et~al.}{1996}]{singh_1996}
{Singh} K.~P.,  {White} N.~E.,   {Drake} S.~A.,  1996, \mn@doi [\apj]
  {10.1086/176695}, \href {http://adsabs.harvard.edu/abs/1996ApJ...456..766S}
  {456, 766}

\bibitem[\protect\citeauthoryear{{Singh} et~al.,}{{Singh}
  et~al.}{2014}]{singh_2014}
{Singh} K.~P.,  et~al., 2014, in Space Telescopes and Instrumentation 2014:
  Ultraviolet to Gamma Ray. p. 91441S, \mn@doi{10.1117/12.2062667}

\bibitem[\protect\citeauthoryear{{Singh} et~al.,}{{Singh}
  et~al.}{2016}]{singh_2016}
{Singh} K.~P.,  et~al., 2016, in Space Telescopes and Instrumentation 2016:
  Ultraviolet to Gamma Ray. p. 99051E, \mn@doi{10.1117/12.2235309}

\bibitem[\protect\citeauthoryear{{Singh} et~al.,}{{Singh}
  et~al.}{2017a}]{singh_2017b}
{Singh} K.~P.,  et~al., 2017a, \mn@doi [Journal of Astrophysics and Astronomy]
  {10.1007/s12036-017-9448-7}, \href
  {https://ui.adsabs.harvard.edu/abs/2017JApA...38...29S} {38, 29}

\bibitem[\protect\citeauthoryear{{Singh}, {Dewangan}, {Chandra},
  {Bhattacharayya}, {Chitnis}, {Stewart}  \& {Westergaard}}{{Singh}
  et~al.}{2017b}]{singh_2017a}
{Singh} K.~P.,  {Dewangan} G.~C.,  {Chandra} S.,  {Bhattacharayya} S.,
  {Chitnis} V.,  {Stewart} G.~C.,   {Westergaard} N.~J.,  2017b, Current
  Science, \href {https://ui.adsabs.harvard.edu/abs/2017CSci..113..587S} {113,
  587}

\bibitem[\protect\citeauthoryear{{Su{\'a}rez Mascare{\~n}o}, {Rebolo}  \&
  {Gonz{\'a}lez Hern{\'a}ndez}}{{Su{\'a}rez Mascare{\~n}o}
  et~al.}{2016}]{suarez_2016}
{Su{\'a}rez Mascare{\~n}o} A.,  {Rebolo} R.,   {Gonz{\'a}lez Hern{\'a}ndez}
  J.~I.,  2016, \mn@doi [\aap] {10.1051/0004-6361/201628586}, \href
  {https://ui.adsabs.harvard.edu/abs/2016A%26A...595A..12S} {595, A12}

\bibitem[\protect\citeauthoryear{{Voges} et~al.,}{{Voges}
  et~al.}{1996}]{voges_1996}
{Voges} W.,  et~al., 1996, \iaucirc, \href
  {https://ui.adsabs.harvard.edu/abs/1996IAUC.6420....2V} {6420, 2}

\bibitem[\protect\citeauthoryear{{Voges} et~al.,}{{Voges}
  et~al.}{2000}]{voges_2000}
{Voges} W.,  et~al., 2000, \iaucirc, \href
  {https://ui.adsabs.harvard.edu/abs/2000IAUC.7432....3V} {7432, 3}

\bibitem[\protect\citeauthoryear{{Wargelin}, {Saar}, {Pojma{\'n}ski}, {Drake}
  \& {Kashyap}}{{Wargelin} et~al.}{2017}]{wargelin_2017}
{Wargelin} B.~J.,  {Saar} S.~H.,  {Pojma{\'n}ski} G.,  {Drake} J.~J.,
  {Kashyap} V.~L.,  2017, \mn@doi [\mnras] {10.1093/mnras/stw2570}, \href
  {http://adsabs.harvard.edu/abs/2017MNRAS.464.3281W} {464, 3281}

\bibitem[\protect\citeauthoryear{Wood \& Linsky}{Wood \&
  Linsky}{2010}]{Wood_2010}
Wood B.~E.,  Linsky J.~L.,  2010, \mn@doi [The Astrophysical Journal]
  {10.1088/0004-637x/717/2/1279}, 717, 1279

\bibitem[\protect\citeauthoryear{{van Leeuwen}}{{van Leeuwen}}{2007}]{van_2007}
{van Leeuwen} F.,  2007, \mn@doi [\aap] {10.1051/0004-6361:20078357}, \href
  {http://adsabs.harvard.edu/abs/2007A%26A...474..653V} {474, 653}

\makeatother
\end{thebibliography}
\end{document}